# UJIAN ONLINE MAHASISWA ILMU KOMPUTER BERBASIS SMARTPHONE

**Leon Andretti Abdillah**
Program Studi Sistem Informasi, Universitas Bina Darma, Palembang
email leon.abdillah@yahoo.com

***Abstract* -** Information technology influence higher education in various aspects, including education sector. This article discusses how smartphones facilitate online examination in computer science and information systems students. The research objective to be achieved by the researchers through the research, are as follows: 1) Utilizing smartphone as a media test online exam, 2) How to make use of social technologies in online test, and 3) Identify the features or facilities that could be used for the implementation of an online exam. Observations was conducted with 87 early year students as respondents. Author develop the online questions by using google forms, and facebook to disseminate online examination questions. Research findings show that Android are dominantly gadgets used by students for their online examination. Smartphone based online exam help students concentration in online exam. Social information technology like facebook and google forms have rich features in supporting online examination for computer science students. The use of smartphones, google forms, and facebook can create an atmosphere of exams modern, efficient, and environmentally friendly.

## 1. PENDAHULUAN

Teknologi informasi (TI) telah menjadi lebih berkolaborasi dengan banyak aspek [1], termasuk dengan perangkat komunikasi, *smartphones*. Perangkat *smartphone* dapat digunakan baik sebagai telepon seluler (*mobile phone*) dan sebagai komputer genggam [2]. Sebagai komputer genggam (*handheld computer*) pribadi, smartphone merupakan langkah terbaru dalam evolusi informasi portabel dan teknologi komunikasi [3], yang mengandung berbagai macam sensor dan komunikasi antarmuka [4]. *Smartphone* sedang diadopsi pada kecepatan yang fenomenal [5] karena *smartphone* memiliki satu set beragam kemampuan *media capture* [6].

*Smartphone* memiliki banyak keuntungan dalam pengiriman suara, teks, gambar, data, dalam format yang kaya dan kecepatan ekstrim. Teknologi dimana-mana (*ubiquitous*) ini dipandang sebagai perangkat serbaguna [7] yang dominan digunakan oleh orang-orang muda.

Terakhir, *smartphone* dapat bertindak teknologi hijau dan sebagai bagian integral dari sistem informasi hijau [6]. Seperti komputer pribadi (PC) atau laptop, *smartphone* juga dioperasikan oleh sistem operasi atau *operating systems* (OS).

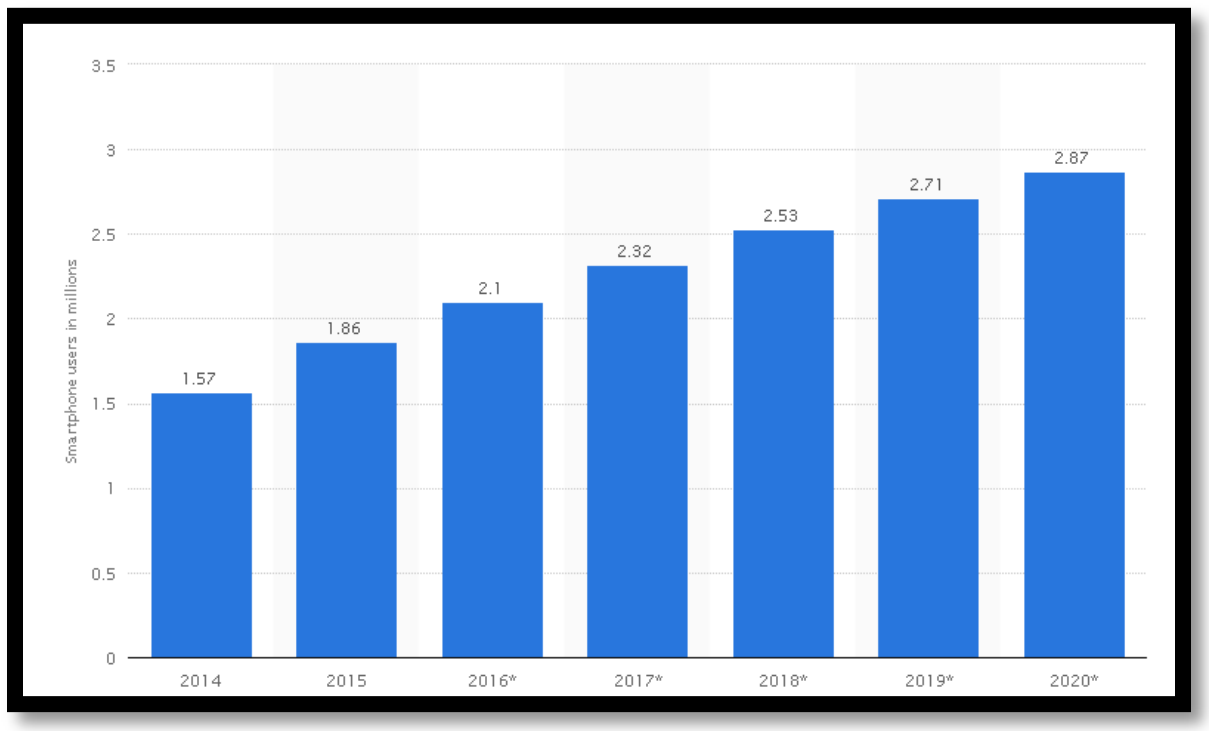

**Gambar 1** Pengguna smartphone di seluruh dunia 2014-2020.

OS merupakan jantung dari sistem perangkat lunak *smartphone* [8]. Hari ini *smartphones* OS didominasi oleh lima pemain utama, yaitu : 1) Android OS - Google Inc., 2) iOS - Apple Inc., 3) Seri 40 (S40) OS - Nokia Inc., 4) BlackBerry OS - BlackBerry Ltd., and 5) Windows OS - Microsoft Corporation. Menurut International Data Corporation (IDC) [9], pada 2016 Q2, Android memimpin pangsa pasar hingga 87,6% diikuti oleh iOS, Windows, dll. (Tabel 1).

**Tabel 1.** Pangsa pasar *Smartphone OS* 2016 Q2.

| OS/Periode | 2015Q3 | 2015Q4 | 2016Q1 | 2016Q2 |
|---|---|---|---|---|
| Android | 84.3% | 79.6% | 83.4% | 87.6% |
| IOS | 13.4% | 18.6% | 15.4% | 11.7% |
| Windows | 1.8% | 1.2% | 0.8% | 0.4% |
| Others | 0.5% | 0.5% | 0.4% | 0.3% |

Sumber: IDC, Agustus 2016

Berdasarkan data pada Tabel 1, Android adalah OS *smartphone* yang utama. Android adalah *platform open source* komprehensif yang dirancang untuk perangkat *mobile* [10] yang terdiri dari paket *software* termasuk OS, *middleware* dan aplikasi inti [11]. Di antara *smartphone* Android [12], Samsung adalah yang paling banyak digunakan oleh vendor (22,8%) diikuti oleh Apple, Huawei, OPPO, Vivo, dll. (Tabel 2).

**Tabel 2.** Pangsa pasar vendor smartphone 2016 Q2.

| OS/Period | 2015Q3 | 2015Q4 | 2016Q1 | 2016Q2 |
|---|---|---|---|---|
| Samsung | 23.3% | 20.4% | 23.8% | 22.8% |
| Apple | 13.4% | 18.6% | 15.4% | 11.7% |
| Huawei | 7.6% | 8.2% | 8.4% | 9.3% |
| OPPO | 3.2% | 3.6% | 5.9% | 1.0% |
| Vivo | 2.9% | 3.0% | 4.4% | 5.9% |
| Others | 49.6% | 46.2% | 42.1% | 40.2% |

Sumber: IDC, Agustus 2016

Pengembangan IT saat ini telah bergabung dengan komunikasi dan pendidikan sektor. Di bidang komunikasi, IT menjadi tulang punggung utama untuk melayani pergerakan data antara *gadgets*. Di bidang pendidikan, IT melayani proses pembelajaran antara dosen dan mahasiswa. Sejak perangkat komunikasi *mobile* yang digunakan di tengah tahun 1990-an, banyak aspek dalam perubahan kehidupan sehari-hari manusia termasuk *ticket reservation* [13], *mobile dictionary* [14], dan *residential locations* [10].

Dalam era akhir-akhir, pendidikan harus mengadopsi dan melibatkan IT dalam proses pembelajaran [15]. *Smartphones* baru dengan semua kemampuannya telah menciptakan lingkungan pembelajaran secara maya atau *virtual learning environment*. *Mobile learning* sebagai lingkungan pembelajaran berbasiskan mobilitas teknologi, mobilitas para pembelajar dan mobilitas pembelajaran yang menambah lanskap pendidikan tinggi [16]. Beberapa penelitian sebelumnya yang terkait dengan *smartphone* dalam pendidikan, sebagai berikut : 1) *Mobile Learning Anytime, Anywhere* [17], 2) Suatu temuan menunjukkan bahwa siswa aktif menggunakan teknologi *mobile* seperti *smartphone* dan komputer *tablet* untuk mendukung pembelajaran mereka [18], dan 3) Siswa percaya perangkat *mobile* yang penting untuk keberhasilan akademis mereka dan menggunakan perangkat mereka untuk kegiatan akademik [19]. Di antara literatur-literatur yang ada, sangat sedikit atau sangat kurang dari mereka yang membahas ujian *online* melalui *smartphone*. Pada saat ujian yang diselenggarakan secara konvensional dilaksanakan masih ditemukan sejumlah masalah, seperti : 1) butuh kertas dan alat tulis, 2) berlangsung secara fisik, sehingga siswa harus berada di tempat yang sama untuk melaksanakan ujian, dan 3) dibutuhkan waktu untuk mengumpulkan lembar jawaban dan proses tabulasi jawabannya. Pada artikel ini, penulis memadukan keunggulan dari *smartphone* dengan media sosial yang paling terkenal, facebook, blog, dan aplikasi google form. Teknologi sosial seperti blog dan media sosial disesuaikan dan ditujukanulang untuk digunakan pendidikan tinggi [20].

Sisa dari artikel ini disusun sebagai berikut. Bagian II berfokus pada bagaimana penelitian dilakukan sebagai metode penelitian. Pada bagian III, penulis menampilkan beberapa informasi diproses atau ditabulasi setelah pemeriksaan dengan menggunakan *smarthphone* sebagai hasil dan diskusi bagian. Akhirnya, peneliti menulis beberapa kesimpulan dan kemungkinan karya masa depan di bagian terakhir, Bagian IV.

## 2. METODE PENELITIAN

### 2.1 Observasi

Peneliti menggunakan observasi kelas untuk menganalisis kegunaan *smartphone* dalam pelaksaan ujian *online* berbasis *smartphone*. Total responden yang terlibat dalam penelitian ini berjumlah mahasiswa baru (mahasiswa tahun awal) yang mengambil kuliah pada bidang program studi sistem informasi (fakultas ilmu komputer). Koleksi data dilakukan ketika siswa melakukan ujian *online* berbasis *smartphone* di kelas dan laboratorium komputer. Posisi duduk mahasiswa ketika uian ditentukan berdasarkan urutan nama mereka dalam daftar kehadiran mahasiswa.

### 2.2 Soal Ujian

Total pertanyaan dalam skema ujian online berjumlah 10 soal. Pertayaan yang diajukan pada ujian online terdiri atas 3 (tiga) tipe soal, yaitu : 1) pertanyaan-pertanyaan pilihan berganda atau *multiple choices* (terdiri atas empat pilihan jawaban yang ditandai dengan pilihan ‘a’ sampai dengan pilihan ‘d’), 2) Pertanyaan-pertanyaan dikotomi atau *dichotomy questions* (menyediakan hanya dua kemungkinan jawaban, *True* atau *False*), dan 3) Pertanyaan-pertanyaan jawaban singkat atau *short answer questions* (mahasiswa diminta mengetikkan jawaban kurang dari tiga kata jawaban). Waktu yang digunakan untuk menyelesaikan ujian *online* adalah selama 75 menit. Sepuluh menit pertama digunakan untuk mengisi *students identifications*. Enam puluh menit selanjutnya dialokasikan untuk menjawab semua pertanyaan ujian *online*. Dan lima menit yang terakhir digunakan untuk *final check* serta *submitting the quizzes*.

## 2.3 *Social Information Technology*

Pada studi ini, para mahasiswa akan menjawab pertanyaan-pertanyaan melalui *link* yang diberikan via media sosial facebook. Facebook digunakan untuk mengelompokkan mahasiswa berdasarkan kelas-kelas mereka. Setelah mahasiswa menge-klik *link* yang telah disediakan, maka google form akan terbuka pada *browser* berisikan pertanyaan-pertanyaan ujian *online*. Google forms bisa digunakan untuk merekam dan menampilkan respon dari para mahasiswa [21]. Apabila pada penelitian-penelitian sebelumnya peneliti memanfaatkan media sosial facebook untuk : 1) *promotion media* [22], 2) *information and knowledge sharing* [23], 3) *political party campaign* [24], dan 4) *presidential social media campaigns* [25]. Maka pada penelitian ini, media sosial facebook akan dimanfaatkan dibidang pendidikan dengan fokus pada dukungan terhadap ujian *online* berbasis *smartphone*.

Dosen membuat lembar soal dengan menggunakan google forms. Setelah semua pertanyaan dengan pilihan jawabannya siap, kemudian dosen mengirimkan *virtual form* melalui : 1) *email*, 2) *Uniform Resource Locator* (URL) atau *shorten URL*, 3) *Embed HTML*. Dosen juga bisa untuk membagikan *online questions forms* ke : 1) Google+, 2) Facebook, atau 3) Twitter.

Dosen perlu melakukan pengaturan (*setting*) untuk mengumpulkan respon mahasiswa melalui google forms. Dosen harus mengatur atau membuat google forms sebagai “quizzes”. Pada pilihan *quiz*, dosen bisa me-*release* nilai secara segera (*immediately*) atau nanti setelah *manual review* (*turns on email collection*). Responden atau mahasiswa dapat melihat : 1) *Missed questions*, 2) *Correct answers*, dan 3) *Point values* (Gambar 2).

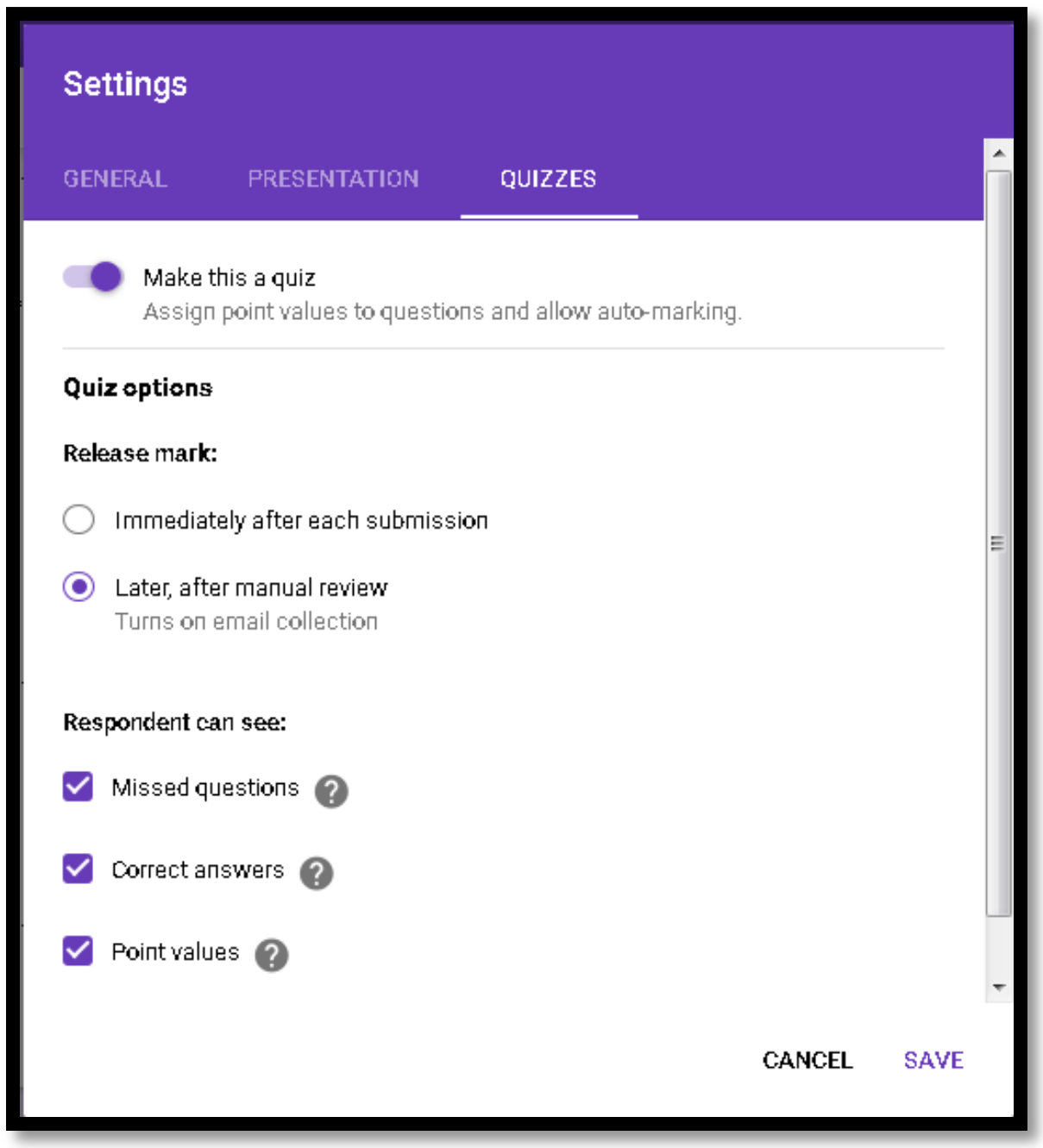

**Gambar 2** Halaman *setting* google forms.

Para mahasiswa akan menjawab semua pertanyaan dari *smartphone* mereka, *personal computer* (PC), atau laptop. Mereka harus memasukkan *google email* yang valid ke google forms. Mengisi semua *required fields* pada google forms. Setelah mahasiswa selesai memasukkan jawaban mereka, suatu respon konfirmasi akan menampilkan suatu pesan konfirmasi. Setiap pertanyaan dalam studi ini memiliki nilai yang sama, yaitu 10 *points*. Google forms dapat menyimpan jawaban yang benar untuk penilaian otomatis.

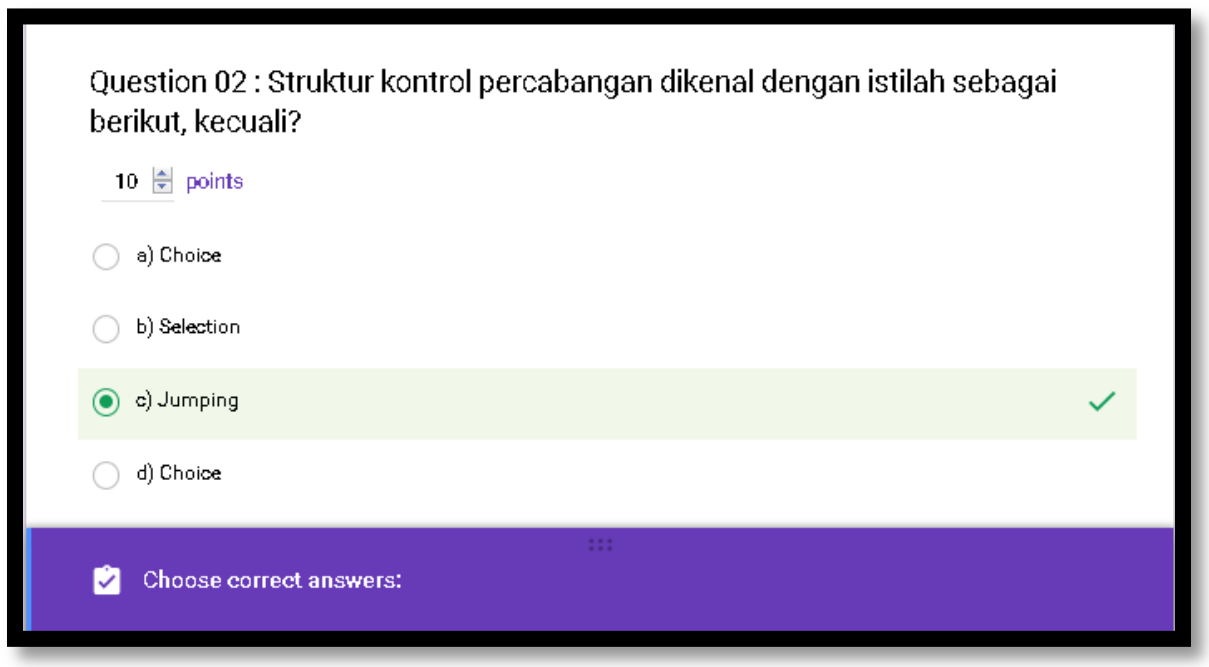

**Gambar 3** *Answer key and points*.

## 3. HASIL DAN PEMBAHASAN

Berdasarkan hasil observasi di kelas, peneliti mampu membuat menintesakan sejumlah *point of views* sebagai hasil. Pada hasil pertama, penulis akan menampilkan suatu *post* di *facebook group*. Post ini digunakan untuk menginformasikan tentang ujian *online* kepada para mahasiswa. Dosen perlu menyediakan tanggal ujian, dan terlebih lagi *time limit*. Dosen juga perlu untuk menyediakan suatu URL yang akan membawa para mahasiswa ke google forms yang berisi soal-soal ujian *online*. Facebook masa kini telah dilengkapi dengan kemampuan menampilkan *preview* dari URL yang terkandung pada suatu *post* (Gambar 4). Setelah mahasiswa menge-klik URL yang diberikan kemudian *browser* akan membawanya ke tautan google forms.

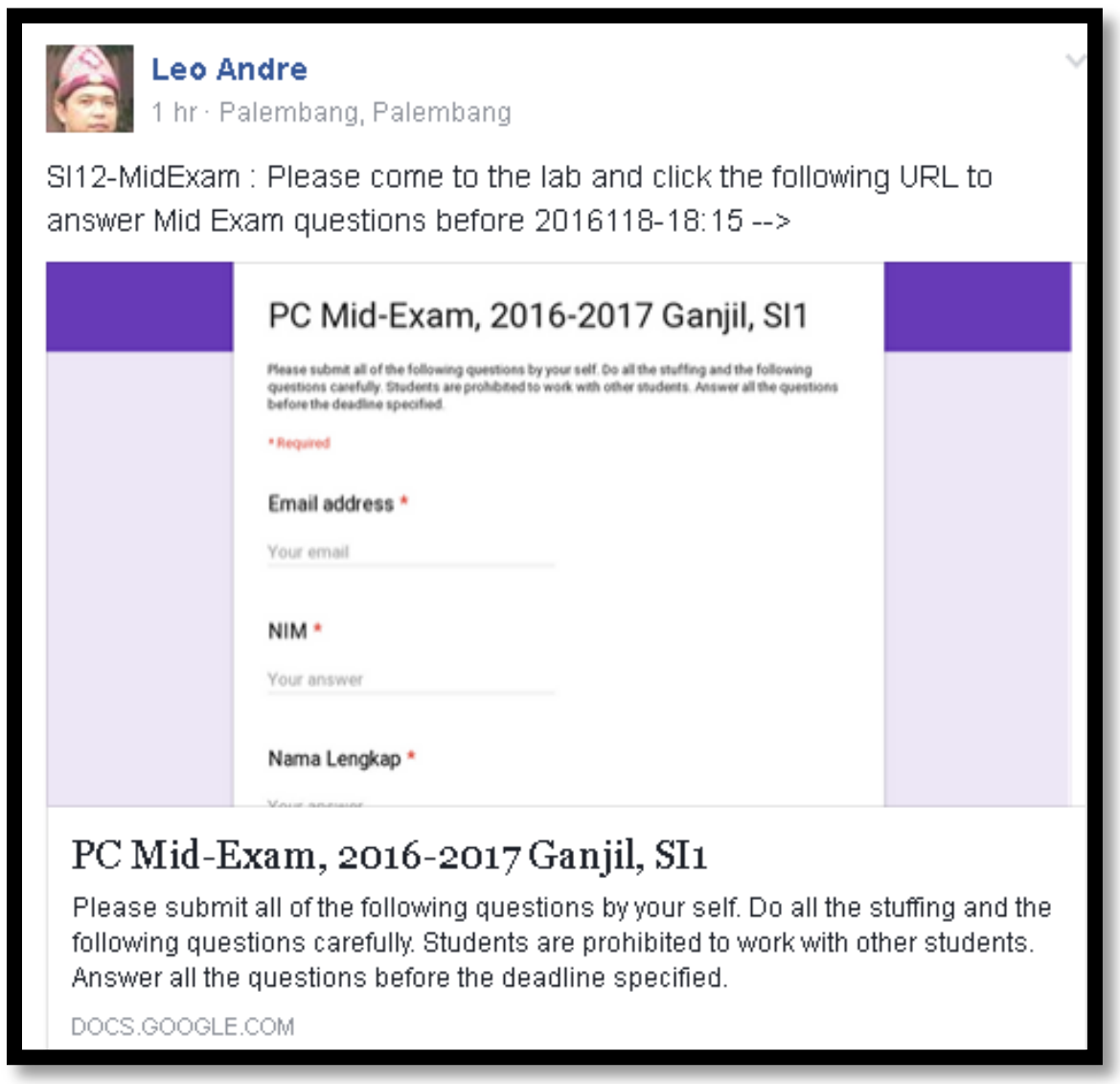

**Gambar 4** Facebook post dari uijan *online*.

Pada virtual google forms ada sejumlah *fields* yang harus diisi oleh para mahasiswa. Untuk studi kali ini, semua *fields* adalah bersifat wajib (*mandatory*). Tanda bintang atau *asterisk* (*) mengindikasikan bahwa *field* tersebut adalah *mandatory*. Tiga *fields mandatory* pertama berupa isian untuk : 1) *email*, 2) nomor induk mahasiswa (NIM), dan 3) Nama lengkap mahasiswa sesuai dengan nama yang tertera pada kartu rencana studi (KRS).

Tipe pertanyaan pertama adalah pertanyaan yang membutuhkan jawaban singkat (*short answers*). Dosen membatasi jawaban yang akan diberikan oleh mahasiswa sebanyak maksimal 2 (dua) kalimat (Gambar 5). Tempat untuk menjawab *short questions* disediakan dibawa pertanyaan yang bersangkutan. Informasi skala skor untuk tiap pertanyaan nampak pada bagian kiri atas seajar dengan pertanyaannya. Tiap pertanyaan memiliki skor sebesar 10 *points*.

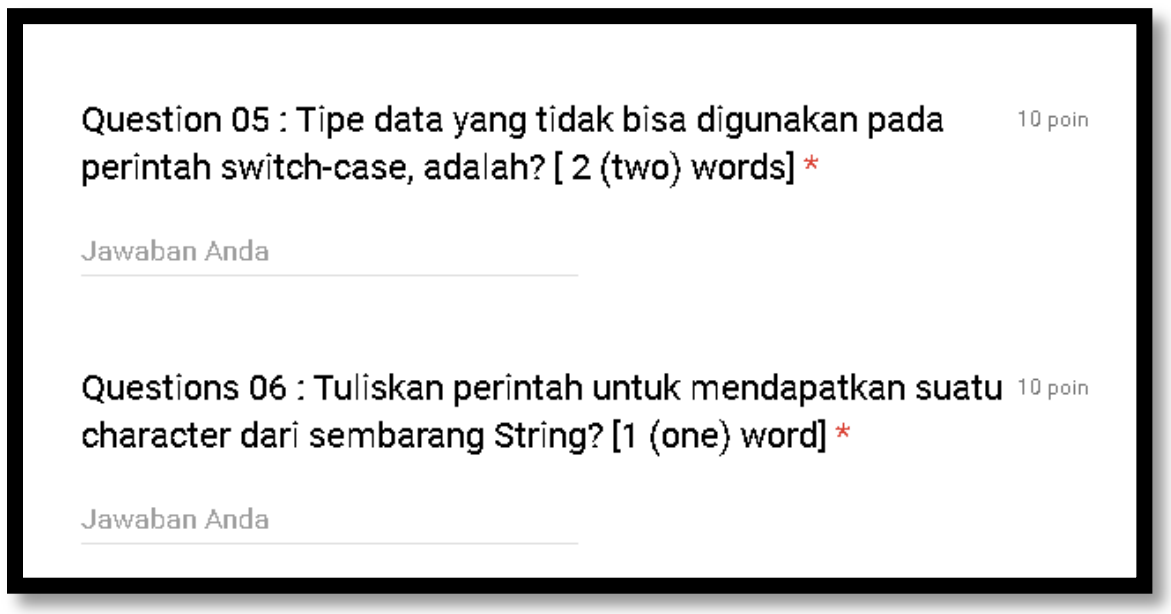

**Gambar 5** Contoh *short questions*.

Tipe pertanyaan kedua adalah *dichotomy questions*. Tipe ini hanya memiliki dua kemungkinan jawaban (*true* atau *false*), lihat Gambar 6 untuk contoh tipe pertanyaan ini. Para mahasiswa hanya perlu menge-klik salah satau pilihan jawaban terbaik untuk tipe soal ini. Tipe pertanyaan *dichotomy* pada dasarnya adalah tipe *multiple-choice questions* yang terdiri atas dua kemungkinan pilihan jawaban.

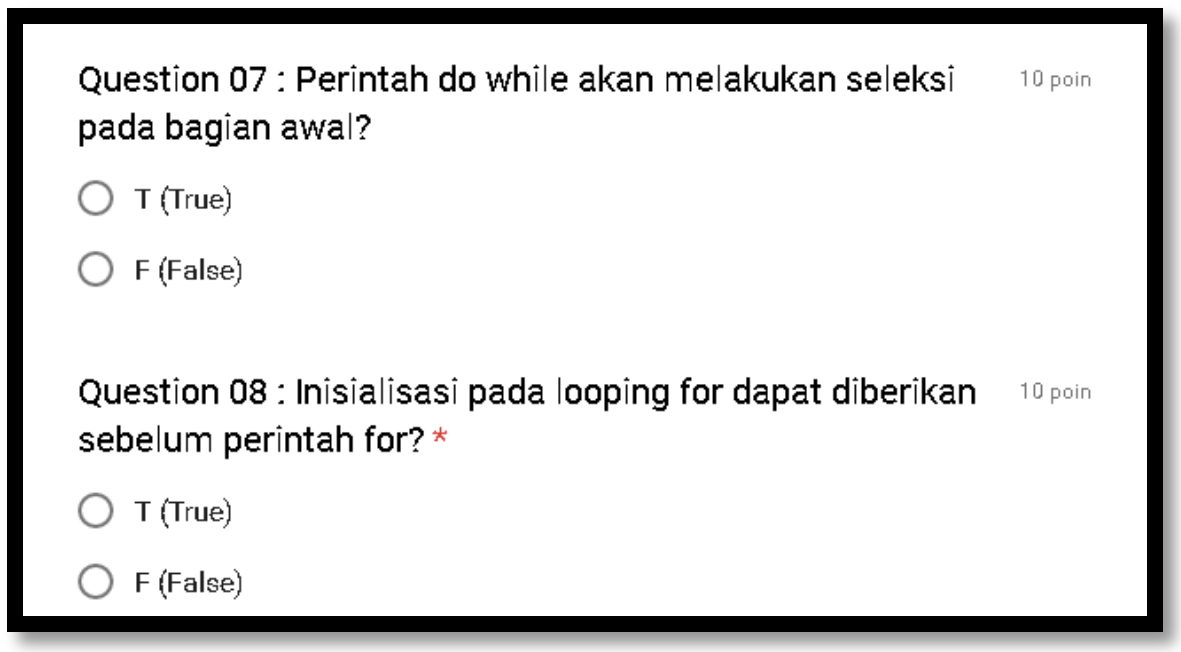

**Gambar 6** Contoh *dichotomy questions*.

Tipe pertanyaan ketiga adalah pilihan berganda atau *multiple choice questions*. Pada tipe ini, ada 4 (empat) kemungkinan pilihan jawabannya. Tiap kemungkinan jawaban diberi kode mulai dari karakter 'a', 'b', 'c' atau 'd' (Gambar 7).

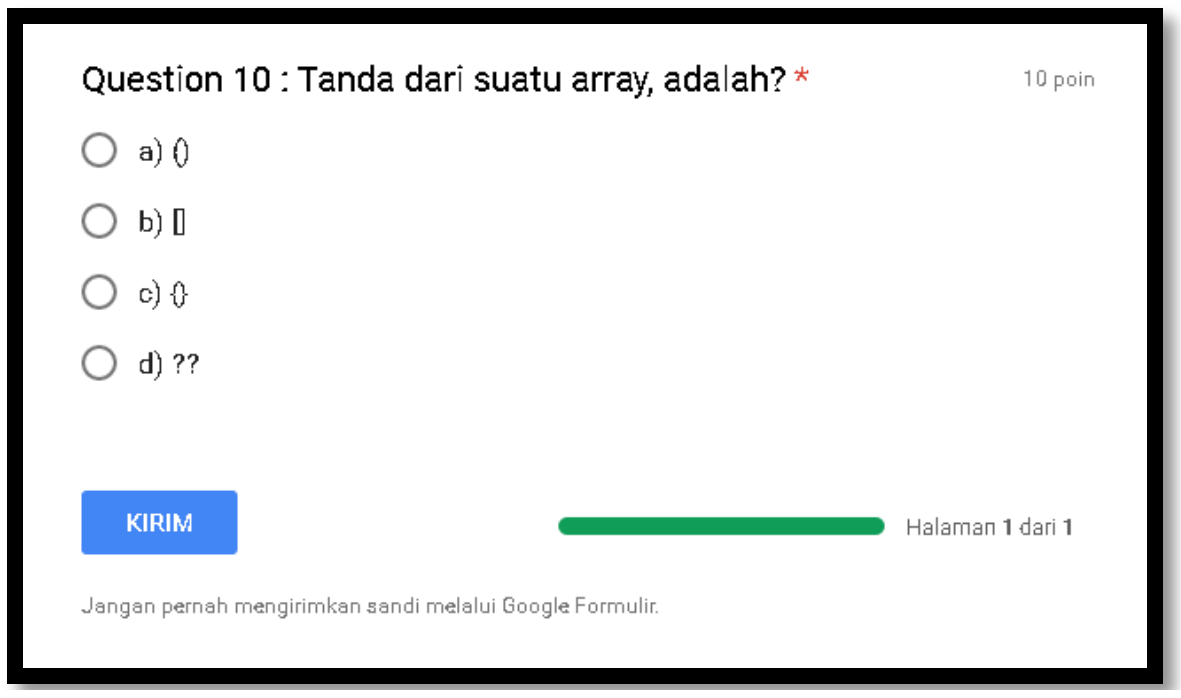

**Gambar 7** Contoh pertanyaan *multiple choice.*

Pada bagian akhir google virtual form ada tombol "Submit" yang memungkinkan para mahasiswa untuk memasukkan semua jawabannya.

### 3.1 Informasi Responden

Jumlah mahasiswa yang terlibat dalam penelitian ini berjumlah 87 orang mahasiswa baru pada fakultas ilmu komputer. Mahasiswa tersebut didominasi oleh mahasiswa laki-laki sebesar 53.9%, sementara mahasiswa perempuan berjumlah 46.1%.

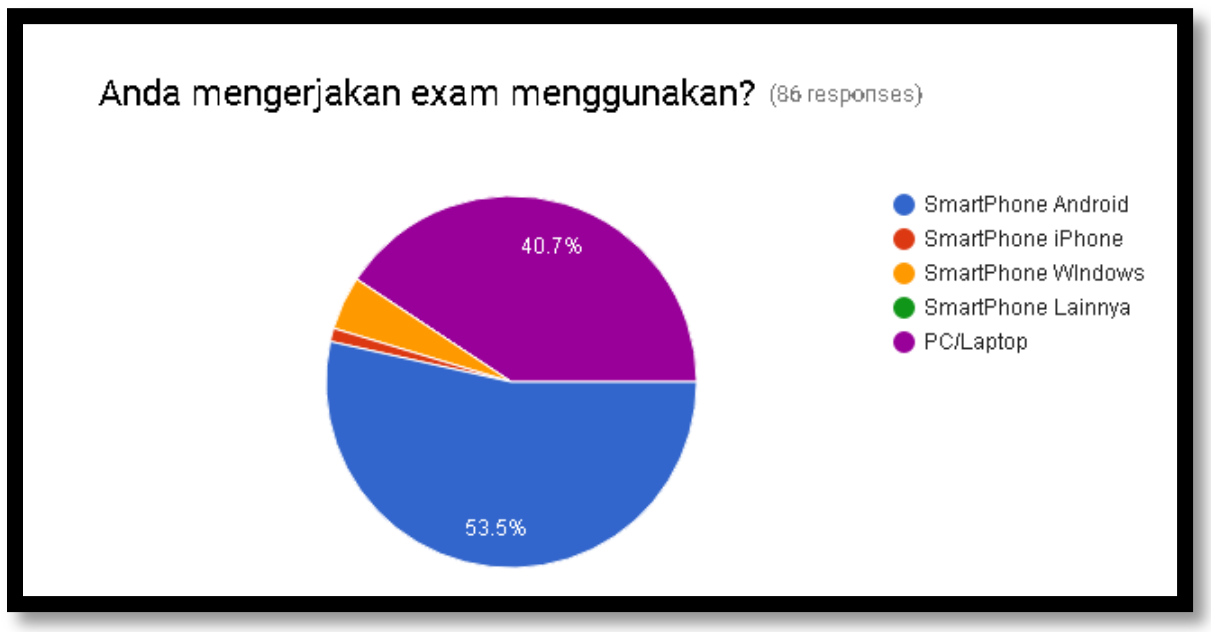

**Gambar 8** Student gadgets.

Walaupun sebagian besar responden merupakan mahasiswa (83.9%), namun beberapa diantaranya telah bekerja. Sekitar 4.7% telah bekerja di perusahaan swasta (*private companies*), 2.3% bekerja sebagai pegawai negeri sipil (PNS) atau perusahaan yang dimiliki oleh pemerintah (BUMN), 1.2% bekera sendiri atau wiraswasta, sedangkan yang menjawab lainnya sebesar 2.3%.

Pada saat ujian *online* berbasis *smartphone* berlangsung, secara keseluruan didominasi oleh mahasiswa pengguna *Android smartphone* yang mencapai 53.5%, diikuti oleh mahasiswa yang menggunakan PC/Laptop mencapai 40.7%, selanjutnya adalah pengguna *smartphone* berbasis Windows sebesar 4.7%, dan seumlah 1.2% dari pengguna iPhone.

**Tabel 3.** Informasi Responden.

| Criteria | Sub Criteria | Percent (%) |
|---|---|---|
| Gender | Men | 53.9% |
| | Women | 46.1% |
| Status | Students | 89.7% |
| | Private company | 4.6% |
| | Civil servants | 2.3% |
| | Self-employed | 1.1% |
| | Other | 2.3% |
| Gadgets | Smartphone Android | 54.0% |
| | PC\|Laptop | 40.2% |
| | Smartphone Windows | 4.6% |
| | Smartphone iPhone | 1.1% |

### 3.2 Konsentrasi Mahasiswa

Berdasarkan hasil pengamatan pada saat ujian *online* berlangsung, para mahasiswa peserta ujian *online* nampak fokus ke layar kecil *smartphone* mereka masing-masing ketika sedang menjawab pertanyaan-pertanyaan ujiannya. Hal ini disebabkan karena layar *smartphone* tidak terlalu besar dibandingkan dengan layar *personal computer* atau laptop. Para mahasiswa juga memiliki keterbatasan untuk memindahkan *page* aktif pada *screen smartphone*-nya.

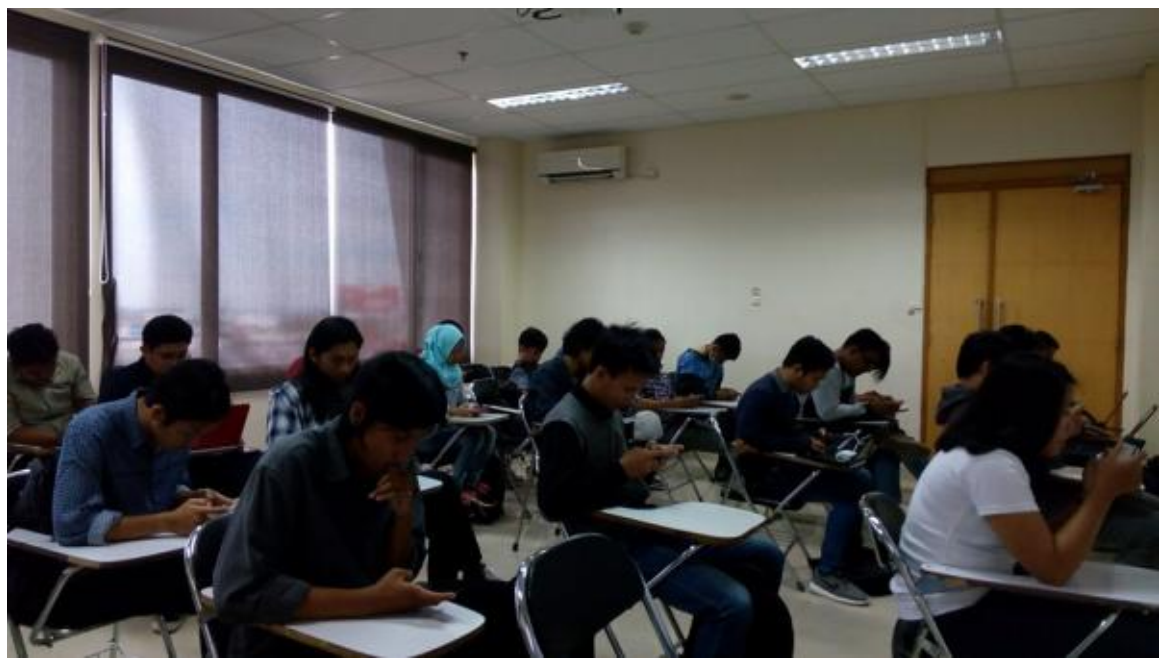

**Gambar 9** Mahasiswa konsentrasi mengerjakan ujian online.

### 3.3 Waktu Respon Nyata (*Real Time*)

Ketika seorang mahasiswa selesai dengan ujian *online*-nya dan kemudian menekan atau menge-klik *submit* dari google form aktifnya, informasi mengenai mahasiswa tersebut akan nampak pada layar dosen selaku *admin*-nya. Dosen dapat melihat siapa mahasiswa yang baru saja mengumpulkan lembar jawaban secara *online* dan *real time* melalui google form. Pada penelitian ini, penulis memantau mahasiswa yang mengumpulkan hasil ujiannya berdasarkan *field* nama mahasiswa.

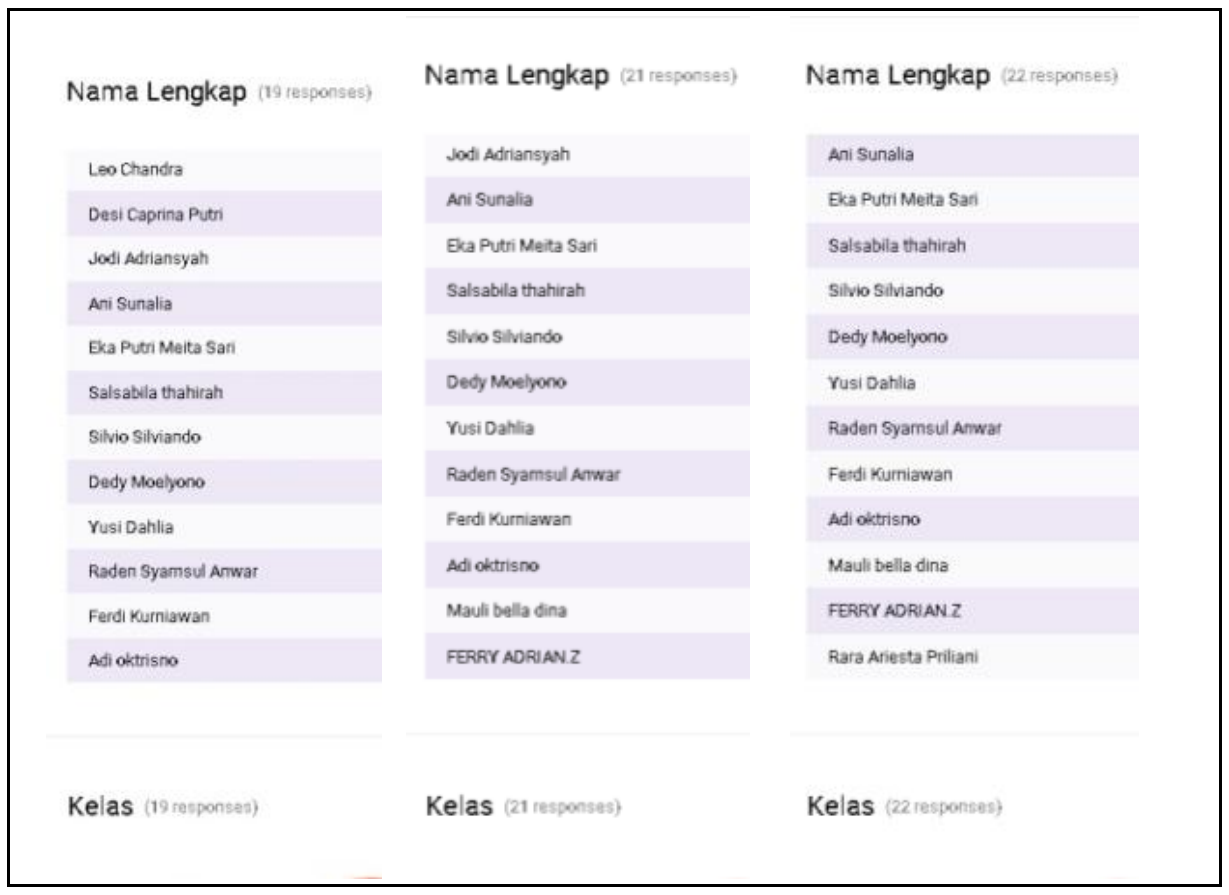

**Gambar 10** Respon real time google forms.

Gambar 10 memperlihatkan bagaimana google forms menampilkan daftar mahasiswa yang baru saja memasukkan jawabannya atau selesai mengerjakan ujian *online*-nya.

### 3.4 Representasi Visual

Setelah semua mahasiswa memasukkan jawabannya melalui google forms dari *smartphone* masing-masing, dosen dapat melihat semua jawaban mahasiswa dalam bentuk tampilan *bar chart*. Salah satu contoh dari hasilnya ditunjukkan pada soal nomor 2 (dua) seperti yang nampak pada Gambar 11.

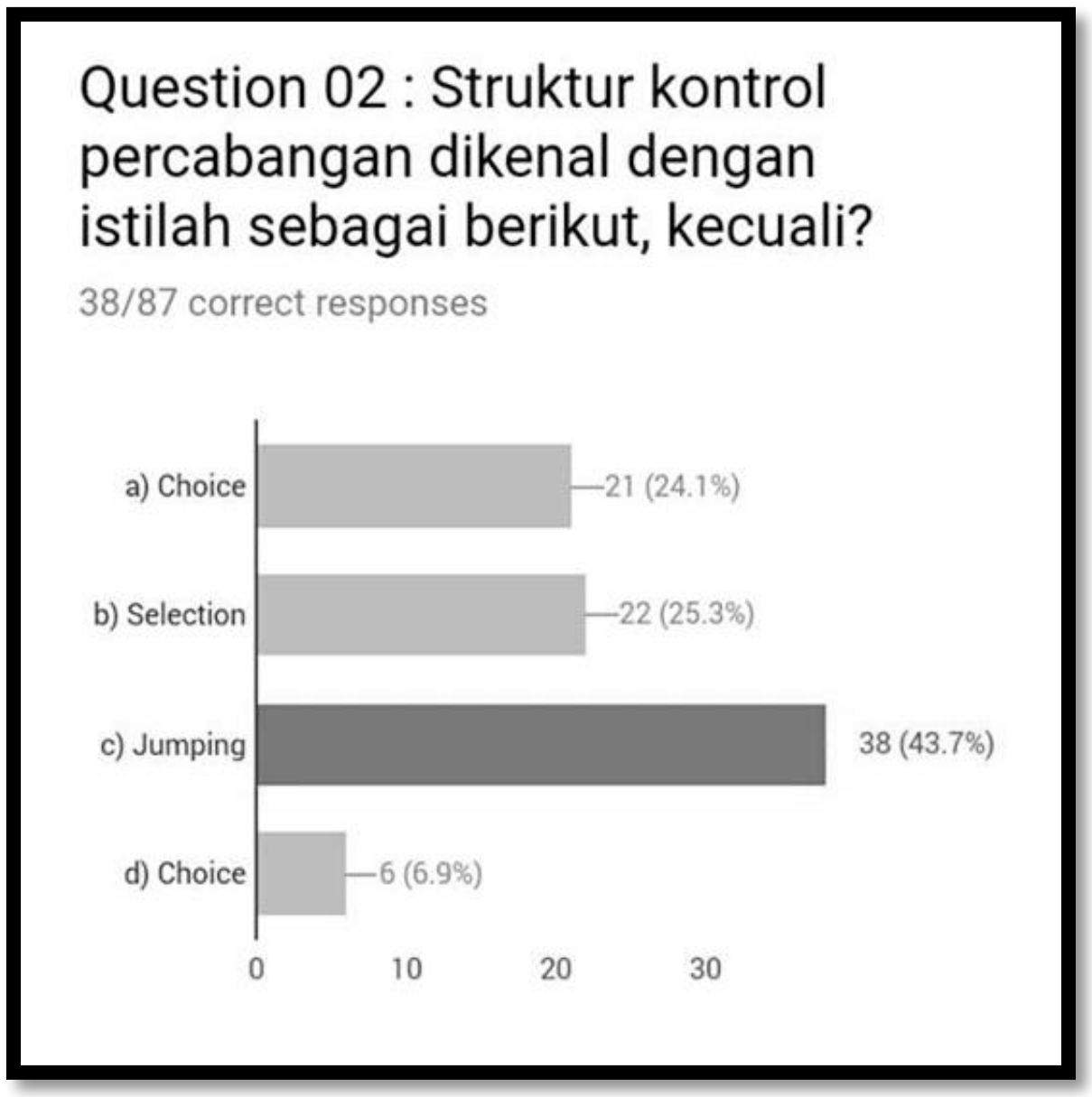

**Gambar 11** Representasi visual google forms.

Gambar 11 menampilkan suatu contoh representasi visual dengan menggunakan bagan batang yang diperkaya dengan persentase. Gambar 10 menginformasikan sekitar 43.7% mahasiswa peserta ujian *online* menjawab dengan benar untuk soal nomor “02”.

Dengan adanya fasilitas ini, maka memudahkan dosen untuk melakukan evaluasi atas penyampaian materi selama perkuliahan. Setiap soal yang diberikan merupakan bagian dari perkuliahan yang dibagi ke dalam sejumlah bab.

### 3.5 Tabulation Cepat

Setelah semua mahasiswa peserta ujian *online* berbasis *smartphone* memasukkan jawaban-nya, google forms menyediakan fasilitas untuk membuka respon jawaban mereka dalam suatu lembar kerja atau *spreadsheet* (*create a new spreadsheet or select existing spreadsheet*). Lembar kerja tersebut mirip dengan lembar kerja Microsoft Excel. Sekali kita memiliki data pada suatu lembar kerja Excel, maka akan dengan mudah untuk diolah sesuai dengan kebutuhan atau rumusan yang digunakan pada penilaian akhir suatu ujian *online* berbasiskan *smartphone* ini.

## 4. KESIMPULAN

Berdasarkan observasi yang dilaporkan pada bagian-bagian sebelumnya, penulis sampai kepada beberapa kesimpulan, sebagai berikut:

1) Ujian *online* berbasiskan *smartpone* memiliki dampak pribadi dalam mempromosikan *green based education*,
2) Sosial teknologi informasi seperti google forms dan facebook memiliki beragam fitur dalam mendukung ujian secara *online*,
3) Ujian *online* berbasiskan *smartphone* mempercepat proses ujian dan pemeriksaan hasil jawabannya, dan
4) Batasan pada penelitian ini adalah terbatasnya cakupan responden.

## REFERENSI